\documentclass[review]{elsarticle}
\usepackage{graphicx}
\usepackage{amsmath}
\usepackage{amssymb}
\usepackage{bm}
\usepackage{subfigure}
\usepackage{color}

\title{GPU-based single-cluster algorithm for the simulation 
of the Ising model} 
\author[tmu]{Yukihiro Komura}
\ead{y-komura@phys.se.tmu.ac.jp}
\author[tmu]{Yutaka Okabe}
\ead{okabe@phys.se.tmu.ac.jp}
\address[tmu]{Department of Physics, Tokyo Metropolitan University, Hachioji, Tokyo 192-0397, Japan}

\begin{document}

\begin{abstract}
We present the GPU calculation with the common unified device architecture 
(CUDA) for the Wolff single-cluster algorithm of the Ising model.  
Proposing an algorithm for a quasi-block synchronization, we realize 
the Wolff single-cluster Monte Carlo simulation with CUDA.  
We perform parallel computations for the newly added spins 
in the growing cluster.
As a result, the GPU calculation speed for the two-dimensional 
Ising model at the critical temperature 
with the linear size $L=4096$ is 5.60 times as fast as the calculation speed 
on a current CPU core.  For the three-dimensional Ising model with 
the linear size $L=256$, the GPU calculation speed is 7.90 times as fast as
the CPU calculation speed. 
The idea of quasi-block synchronization can be used not only 
in the cluster algorithm but also in many fields where the synchronization 
of all threads is required. 
\end{abstract}


\maketitle
\section{Introduction}
The Ising model is a simple and standard model of statistical physics. 
The two-dimensional (2D) Ising model, which was exactly solved 
by Onsager \cite{onsager44}, shows a phase transition 
at the critical temperature $T_{c}/J= 2.269 \cdots$. 
The Ising model has been studied by using many Monte Carlo techniques, 
including the Metropolis algorithm \cite{metro53}, cluster algorithm 
\cite{sw87,wolff89}, extended ensemble algorithm such as 
the Wang-Landau method \cite{wl01}. 
Since the Ising model is simple and its exact solution is available, 
it is sometimes used to test a new Monte Carlo algorithm.

Recently the speed-up of computation with graphics processing unit (GPU) 
has captured a lot of attention in many fields including 
computational biology and chemistry 
\cite{biology_chemistry_cite_No1}, 
molecular dynamics simulation of fluids \cite{fluid_dynamics_standart, fluid_dynamics_new}, 
CT image reconstruction \cite{GPU_CT_image_09,GPU_CT_image_cite_No1}, 
finance \cite{preis_finance 09} and much more. 
GPU has been developed as a supplementary arithmetic device of 
image processing.  GPU has many streaming multiprocessors (SM). 
Each SM is composed of eight streaming processors (SP) with 
a multi-threaded instruction unit and shared memory. 
GPU has realized the acceleration of image processing by using many SMs 
in parallel.  
Since GPU is specialized for highly parallel computation, 
it is designed such that many transistors are devoted to data processing. 
The common unified device architecture (CUDA) released by NVIDIA 
makes it possible to make a general purpose computing on GPU. 
CUDA allows us to implement algorithms using standard C language 
with CUDA specific extensions. 
The host program launches a sequence of kernels. 
A kernel is organized as a hierarchy of threads. 
Threads are grouped into blocks, and blocks are grouped into a grid. 
The organization of a grid is determined when the GPU 
"kernel" function is invoked as 
\begin{equation}
   kernel\_ function <<< grid, block>>> (arguments); 
\end{equation}
Here, $grid$ and $block$ are integer variables, 
which specify the number of blocks and that of threads, respectively.
For simplicity, we have used one-dimensional representation 
for the dimensions of $grid$ and $block$. 
Threads in a single block will be executed on a single SM, 
sharing the data cache, and can synchronize and 
share data with threads in the same block. 
There are several types of memories, and 
there is a significant difference in the access speed of memories. 
The access speed of global memory, which is accessible 
from either all the threads or the host, is much slower 
than that of shared memory, which is accessible 
from the threads in one block.
The access speed of register, which is only accessible by the thread, 
is nearly equal to that of shared memory. 

Preis {\it et al.} \cite{preis09} studied the 2D and 
three-dimensional (3D) Ising model 
by using Metropolis algorithm with CUDA.  They used a variant of sublattice 
decomposition for a parallel computation on GPU. 
The spins on one sublattice do not interact with other spins 
on the same sublattice.  Therefore one can update all spins 
on a sublattice in parallel when making the Metropolis simulation. 
As a result they were able to accelerate 60 times for the 2D Ising model 
and 35 times for the 3D Ising model compared to a current CPU core. 
More recently, the GPU acceleration of the multispin coding 
of the Ising model was reported \cite{block10}.

Metropolis algorithm \cite{metro53} has flexibility which allows 
its application to a great variety of physical systems. 
However, this algorithm is sometimes plagued by long autocorrelation time, 
named critical slowing down. 
The cluster update algorithms are efficient for overcoming the problem of 
critical slowing down. 
The autocorrelation time drastically decreases by using the Wolff 
single-cluster algorithm \cite{wolff89}. 
It is highly desirable to apply parallel computations to cluster 
algorithms.  But only a limited number of attempts have been reported 
so far along this line.  
The message passing interface (MPI) parallelization technique 
was employed by Bae {\it et al.} \cite{bae} for the Wolff algorithm. 
Quite recently, Kaupuzs {\it et al.} \cite{kaupuzs} studied 
the parallelization of the Wolff algorithm by using 
the open multiprocessing (OpenMP). 
For the 3D Ising model with linear size $L$=1024, 
they reached the speed-up about 1.79 times on two processors 
and about 2.67 times on four processors, as compared to the serial code.

In this paper, we present the GPU calculation with CUDA for the Wolff single-cluster algorithm of the Ising model. 
The rest of the paper is organized as follows. 
In section 2, we briefly describe the standard way of implementing 
Wolff single-cluster algorithm on a CPU. 
In section 3, we explain the idea of quasi-block synchronization on GPU. 
In section 4, we show the idea and techniques of GPU calculation 
for the Wolff single-cluster algorithm. 
In section 5, we compare the performance of GPU calculation 
with that of CPU calculation. 
The summary and discussion are given in section 6.

\section{Wolff single-cluster algorithm}
Our Hamiltonian is given by
\begin{equation}
 \mathcal{H} = -J\sum_{<i,j>}S_{i}S_{j}.
\end{equation}
Here, $J$ is the coupling and $S_{i}$ is the Ising spin on the lattice site $i$. 
The summation is taken over the nearest neighbor pairs $<i,j>$. Periodic boundary conditions are employed. 

Wolff proposed a Monte Carlo algorithm in which only a single cluster 
is flipped at a time \cite{wolff89}. 
The spin-update process of the Wolff single-cluster algorithm 
on a CPU can be formulated as follows \cite{janke,landau}: 
\begin{itemize}
\item[(i)] Choose a site $i$ randomly. 
\item[(ii)] Look at each of the nearest neighbors $j$. If $S_j$ 
is parallel with $S_i$, add $S_j$ to the cluster 
with probability $p=1-e^{-2\beta}$, where $\beta$ is 
the inverse temperature $J/T$. 
Then, flip $S_i$.
\item[(iii)] 
If all the nearest neighbors $j$ have been checked, 
look at each of the nearest neighbors $k$ of site $j$. 
If $S_k$ is parallel with $S_j$ and it was not a member of 
the cluster before, add $S_k$ to the cluster 
with probability $p=1-e^{-2\beta}$.  
Then, flip $S_j$.
\item[(iv)] For all the added spins $S_{k}$'s, 
repeat step (iii) until no more new bonds are created.
\item[(v)] Go to (i).
\end{itemize}
Slightly different ways of formulation are possible, for example, 
for the timing of the flip of spins.
To check the condition that no more new bonds are created in (iv), we may 
use two variables \verb+ic+ and \verb+in+, where \verb+ic+ is the total number 
of members of cluster, and \verb+in+ is the total number of sites 
which have been already checked.  We repeat the step (iii) 
while the condition $\verb+in+ < \verb+ic+$ is satisfied.
We note that the check of site $j$ is done sequentially 
on a standard CPU. 

\section{Quasi-block synchronization}
In parallel computing, the best performance is obtained if calculation 
on each thread is done independently.  At the same time, 
barrier synchronization is needed at a certain point.  
CUDA provides a barrier synchronization 
function \verb+syncthreads()+.  When a kernel function calls 
\verb+syncthreads()+, all threads in a block will be held 
at the calling location until everyone else in the block 
reaches the location.  
But CUDA is not equipped with an inter-block synchronization function. 
Since the synchronization for different blocks is required 
in our GPU calculation of cluster algorithm, 
we here propose an idea of a quasi-block synchronization. 

In CUDA, the organization of a grid is determined 
by two special parameters provided during kernel launch 
as mentioned in section 1.  They are the size of the grid 
in terms of numbers of blocks, $grid$, and the size of 
each block in terms of numbers of threads, $block$.  
Moreover, all the threads are identified by two-level unique 
coordinates, called \verb+blockIdx.x+ and \verb+threadIdx.x+.
We note that \verb+blockIdx.x+ is the number associated 
with each block within a grid, and \verb+threadIdx.x+ 
refers to the label associated with a thread in a block. 
Here, we have used one-dimensional IDs. 

Now, we present an idea of synchronization of threads 
across the blocks.  
The program of quasi-block synchronization is as follows: 
\begin{verbatim}
 /*    quasi-block synchronization    */
   if(threadIdx.x == 0){ stopper[blockIdx.x] = n; }      (a)
   if(threadIdx.x == 0){ stop = 0; }                     (b)
   __syncthreads();
   while(stop != 1){
      if(threadIdx.x == 0){ stop = 1; }                  (c)
      __syncthreads();
      if(stopper[threadIdx.x] < n){ stop = 0; }          (d)
      __syncthreads();
   }                      // while(stop != 1) loop end
\end{verbatim}
We use a flag variable \verb+stop+, which is allocated in a shared memory. 
Only after \verb+stop+ of every block becomes 1, one gets out of 
the "while" loop of quasi-block synchronization. 
To check whether each block reaches this location, we use the array 
\verb+stopper[ ]+, which is allocated in a global memory.  
As an initial value, we set \verb+stopper[blockIdx.x]+ as $m$, 
which is smaller than $n$, for all blocks. 
When any block reaches this location, \verb+stopper[blockIdx.x]+ 
is set as $n$ in (a). 
If there are \verb+stopper[blockIdx.x]+ which are smaller than $n$, 
then \verb+stop+ will be set as 0 for all the blocks in (d).
The trick is that one sets \verb+stopper[blockIdx.x]+ in (a) 
but one checks \verb+stopper[threadIdx.x]+ in (d).
It is natural that we choose the size of grid, $grid$, 
and the size of block, $block$, as the same value. 
In this way, we can realize the inter-block synchronization. 
Of course, the synchronization within a block is guaranteed 
by the synchronization function, \verb+syncthreads()+.

When we execute the quasi-block synchronization consecutively, 
we use different values for $n$ in (a).  
There is a possibility that 
some block reaches the next gate before some other block 
gets out of this routine completely. 
We escape from this problem by increasing $n$ by one. 
To make a loop for $n$ between $n_{\rm min}$ and $n_{\rm max}$, 
we slightly modify the condition (d) at the points of 
$n_{\rm min}$ and $n_{\rm max}$.

We here mention the choice of the size of grid and that of block.
To design the organization of a grid, we take account of 
the size of shared memory and register, the number of processors 
which are executed simultaneously within a block, etc. 
A warp in CUDA is a group of 32 threads, which is the minimum size 
of the data processed in SIMD  (Single Instruction Multiple Data) 
fashion.  Thus, we choose $grid$ and $block$ as 32 
in our GPU calculation of cluster algorithm.  
Then, the total number of threads in a grid 
becomes 1024 ($= 32 \times 32$).  
Since the maximum number of threads per block is 512 
for NVIDIA GeForce GTX 200 series, we could use a single block. 
However, it is not efficient because the number of 
processors executed simultaneously in a block is limited. 

Although it is natural to choose $grid$ and $block$ as the same value, 
we can remove this restriction for $block > grid$ 
by imposing a condition such that 
\begin{verbatim}
      if(threadIdx.x < grid)
         if(stopper[threadIdx.x] < n){ stop = 0; }       (d')
      } 
\end{verbatim}
in (d).
We should note that there are upper limits for the choice 
of $grid$, the total number of blocks, and $block$, the total number 
of threads.  The upper limit of $block$ is determined by 
the maximum number of threads per block, and that of $grid$ 
by the number of blocks in one SM and the number of SMs. 
The number of blocks in one SM is determined by 
the number of threads and the amount of registers and shared memories, 
and the number of SMs depends on the model of GPU.

We may employ other methods for inter-block synchronization.  
One candidate is to use atomic operations such as atomicAdd() 
provided by CUDA.  Atomic operations are performed without interference 
from any other threads.  Inter-block synchronization is realized 
by checking that a flag variable in global memory is accessed 
from all the blocks. However, it takes time to use atomicAdd() 
when many blocks 
try to access to the global memory for flag variable at the same time.
Xiao and Feng \cite{xiao} proposed two approaches 
for inter-block GPU synchronization.  They compared the GPU lock-based 
synchronization using atomic function with the GPU lock-free synchronization. 
Our algorithm is a refined version of the latter, the GPU lock-free 
synchronization, in the sense that the number of access to 
global memories are reduced. 
They found that the GPU lock-based synchronization takes more time 
than the GPU lock-free synchronization.
Another possible method is the master-slave method 
\cite{master_slave_global_barrier} by Volkov and Demmel.
The time of global memory accesses should be carefully checked. 
Further study on inter-block synchronization will be required.

\section{GPU calculation of the Wolff single-cluster algorithm}

In this section, we describe the GPU calculation for the Wolff 
single-cluster algorithm.  
The sublattice decomposition cannot be used for parallelization 
in the case of cluster algorithm. 
We here perform parallel computation for the newly added spins $S_k$ 
in the step (iii) of section 2. 
This idea is similar to that by Kaupuzs {\it et al.} \cite{kaupuzs}, 
where these newly added spins in the wave front of the 
growing cluster were referred to as wave-front spins.
We assign each of wave-front spins to the thread in the grid. 

There are several points which ensure that the parallel algorithm 
works correctly.  We should avoid the situation where different 
threads try to incorporate the same spin to the cluster simultaneously. 
To do this, after checking one direction of nearest neighbors, 
we perform both thread synchronization and quasi-block 
synchronization.  Moreover, the newly added spins 
should be numbered.  In such a numbering, the synchronization 
after the process of each direction of neighbors is essential. 
The numbering and the update of \verb+ic+, the total number 
of members of cluster, are done 
when the last direction of neighbors is checked for each spin.

Then, the step (iii) is modified as follows:
\begin{itemize}
\item[(iii')] All the processes are done for each thread 
representing the wave-front spin. 
After checking each direction of nearest neighbors, 
perform the synchronization.  
When the last direction of the neighbors is checked, 
sum up the total number of newly added spins, 
and complete the numbering of the newly added spins. 
\end{itemize}

We here make a remark on the implementation of parallel computation. 
The number of wave-front spins is represented by $\verb+(ic-in)+$ 
in terms of \verb+ic+ and \verb+in+ discussed in section 2. 
If the thread is indexed by 
\begin{verbatim}
    index = blockIdx.x * 32 + threadIdx.x;
\end{verbatim}
and the kernel function is invoked on the condition that
$\verb+index+ < \verb+(ic-in)+$,
we make calculations only for wave-front spins.

Then, the main part of GPU kernel function is designed as 
\begin{verbatim}
    if( index < (ic-in) ){
        "check one direction of the neighbors of wave-front spin"
    }
    "perform synchronization"
    ...
    ...
    if( index < (ic-in) ){
        "check the last direction of the neighbors of wave-front spin"
    }
    "perform synchronization"
    "flip the wave-front spin"
    "update the numbering of the newly added spins and ic"
\end{verbatim}

As mentioned before, we chose the size of the grid in terms of threads as 1024
$( = 32 \times 32)$.  Then, actually, the upper limit of 
the parallelization is 1024.  In other words, the calculation 
is made in parallel for min($\verb+(ic-in)+$,1024). 
Thus, \verb+in+, the total number of sites which have been 
already checked, is updated as $\verb"in"$ = 
$\verb"in" + \min(\verb"(ic-in)",1024)$.

In making the GPU calculation with CUDA, the proper use of 
shared memories, the technique to avoid the warp divergence 
or the bank conflict when taking the summation 
over the shared memories, etc., are very effective 
for fast computation. 
Since the access to global memory is time-consuming, 
it is better to reduce the frequency of the access to global memory. 
For this purpose we use the following data structure; 
the information on spin value and that on flag index 
to specify whether the site is a member of the cluster or not 
are put into one word.

We finally note that we use a linear congruential random generator 
which was proposed by Preis {\it et al.} \cite{preis09} 
when evaluating the transition probability $p=1-e^{-2\beta}$.

\section{Results}
We have tested the performance of our code on NVIDIA GeForce GTX 285. 
For comparison, we run the code on a current CPU, 
Intel(R) Xeon(R) CPU W3520 @ 2.67GHz.  Only one core of the CPU 
is used.  For compiler, we have used gcc 4.2.1 with option -O3. 

Since the cluster size is dependent on the temperature, 
the computational time changes with temperature. 
So we compare the GPU computational time with the CPU computational time 
at the critical temperature, 
$T_{c}/J = 2.269$ for the 2D Ising model and $T_{c}/J = 4.510$ 
for the 3D Ising model. 
The average computational times per a single-cluster update 
at the critical temperature for the 2D and 3D Ising model 
are tabulated in table \ref{tb:GPU_CPU_time_2D} and 
table \ref{tb:GPU_CPU_time_3D}, respectively.  There, the time 
for only spin-update and 
that including the measurement of energy and magnetization are given.
We show the measured time in units of micro sec. 
The linear system sizes ($L=N_x=N_y=N_z$) are $L=512, 1024, 2048, 4096$ 
for the 2D Ising model 
and $L=64, 128, 256$ for the 3D Ising model. 
We can see from tables \ref{tb:GPU_CPU_time_2D} and \ref{tb:GPU_CPU_time_3D} 
that the acceleration of our single-cluster algorithm 
increases as the linear system size grows to a large size. 
The GPU computational speed for the 2D Ising model with $L=4096$ 
is 5.60 times as fast as the CPU computational speed, 
and that for the 3D Ising model with $L=256$ is 
7.90 times as fast as the CPU computational speed. 
This result indicates that the acceleration rate of 
our single-cluster algorithm surpasses that of the algorithm by 
Kaupuzs {\it et al.} \cite{kaupuzs}, which reached the speed-up 
about 2.67 times for the 3D Ising model with linear size $L$=1024. 
Because of the size of memory for GPU, that is, 1GB 
for NVIDIA GeForce GTX 285, the system size is currently limited 
to $L=4096$ for 2D and $L=256$ for 3D.  With the increase of 
the system size available, much better performance is expected 
for $L$=1024 in 3D.

\begin{table*}[htbp]
\begin{center}
\begin{tabular}{rrrrr}
\hline
$L$ \quad & GPU \quad\quad    &                     & CPU \quad\quad    & \\
          & \quad update only & \quad + measurement & \quad update only &
\quad + measurement\\
\hline
  512 & 7.634 msec & 7.677 msec& 4.151 msec& 4.913 msec \\
 1024 & 15.80 msec & 16.10 msec& 20.97 msec& 24.19 msec \\
 2048 & 33.02 msec & 34.50 msec& 87.59 msec& 100.9 msec \\
 4096 & 81.37 msec & 83.95 msec& 412.0 msec& 470.9 msec \\
\hline
\end{tabular}
\caption{\label{tb:GPU_CPU_time_2D}Average computational time per a single-cluster update 
at $T_c$ for the 2D Ising model.  The time for only single-cluster 
update and that including the measurement of energy and magnetization are given.}
\end{center}
\end{table*}

\begin{table*}[!htbp]
\begin{center}

\begin{tabular}{rrrrr}
\hline
$L$ \quad & GPU \quad\quad    &                     & CPU \quad\quad    & \\
          & \quad update only & \quad + measurement & \quad update only &
\quad + measurement\\
\hline
  64 & 1.176 msec & 1.378 msec& 1.170 msec& 1.947 msec \\
 128 & 2.801 msec & 3.483 msec& 9.785 msec& 15.76 msec \\
 256 & 13.04 msec & 17.45 msec& 97.47 msec& 137.9 msec \\
\hline
\end{tabular}
\caption{\label{tb:GPU_CPU_time_3D}Average computational time per a single-cluster update 
at $T_c$ for the 3D Ising model.  The time for only single-cluster 
update and that including the measurement of energy and magnetization are given.}
\end{center}
\end{table*}

Next, we refer to the temperature dependence of our single-cluster algorithm. 
We plot the temperature dependence of the acceleration rate, that is, 
the ratio of the CPU computational time to the GPU computational time 
for the 2D Ising model with $L=4096$ and 
that for the 3D Ising model with $L=256$ in figures 1(a) and (b), respectively. 

The time for only spin-update and that including the measurement of energy and 
magnetization are shown there. 
We notice from figures 1(a) and (b) that 
the acceleration of computational speed increases at the temperatures 
apart from the critical temperature. 
The reason of accelerating the computational speed at low temperature is 
that the cluster size grows larger 
and the parallel computing becomes more effective. 
On the other hand, the cluster size becomes smaller at high temperature, but 
the efficiency of measurement of energy and magnetization in parallel computing becomes prominent. 
To conclude, our single-cluster algorithm is effective 
over a wide range of temperatures.  

\begin{figure}
\begin{center}
\includegraphics[width=0.4\linewidth]{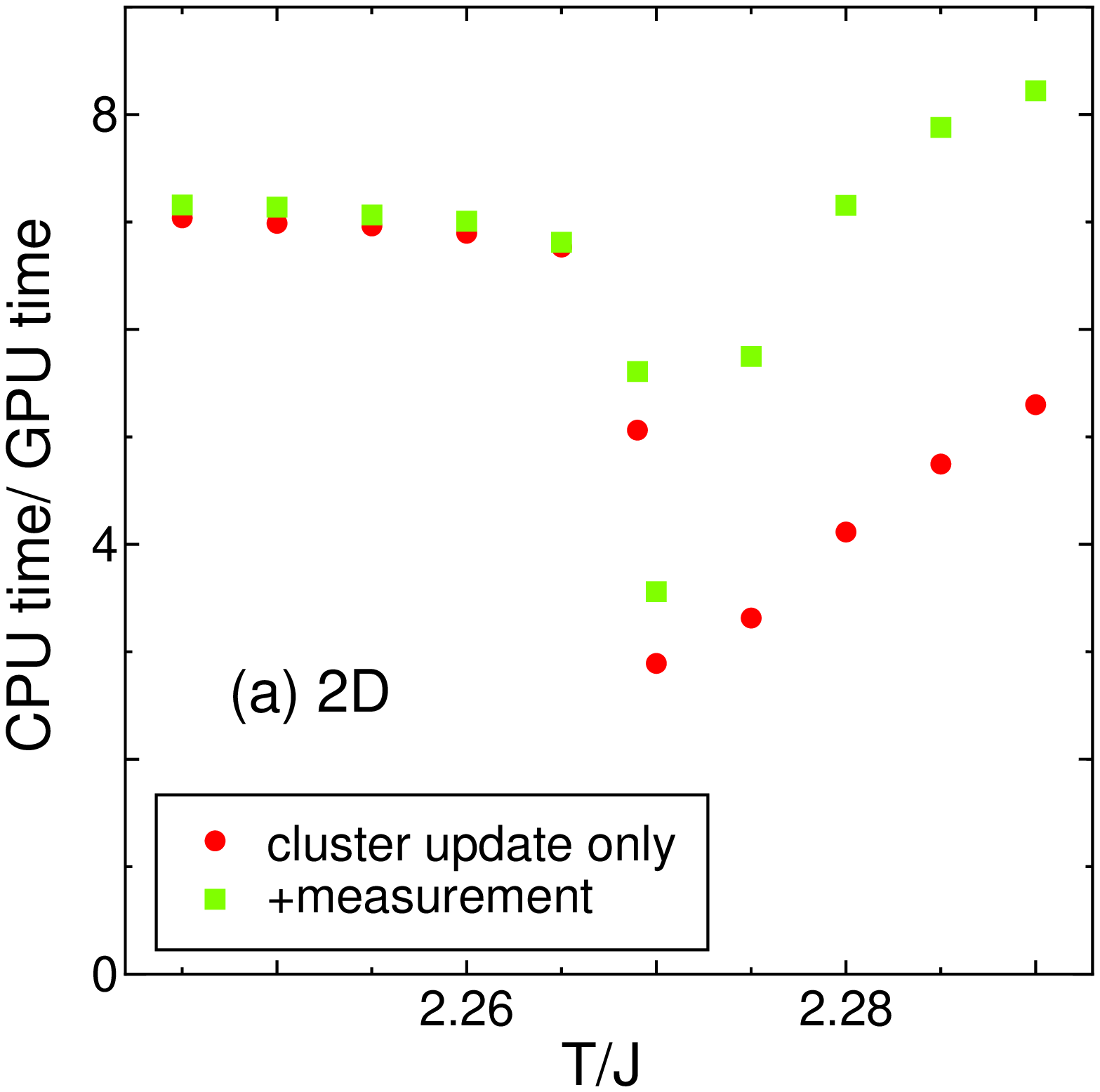}
\hspace{0.02\linewidth}
\includegraphics[width=0.4\linewidth]{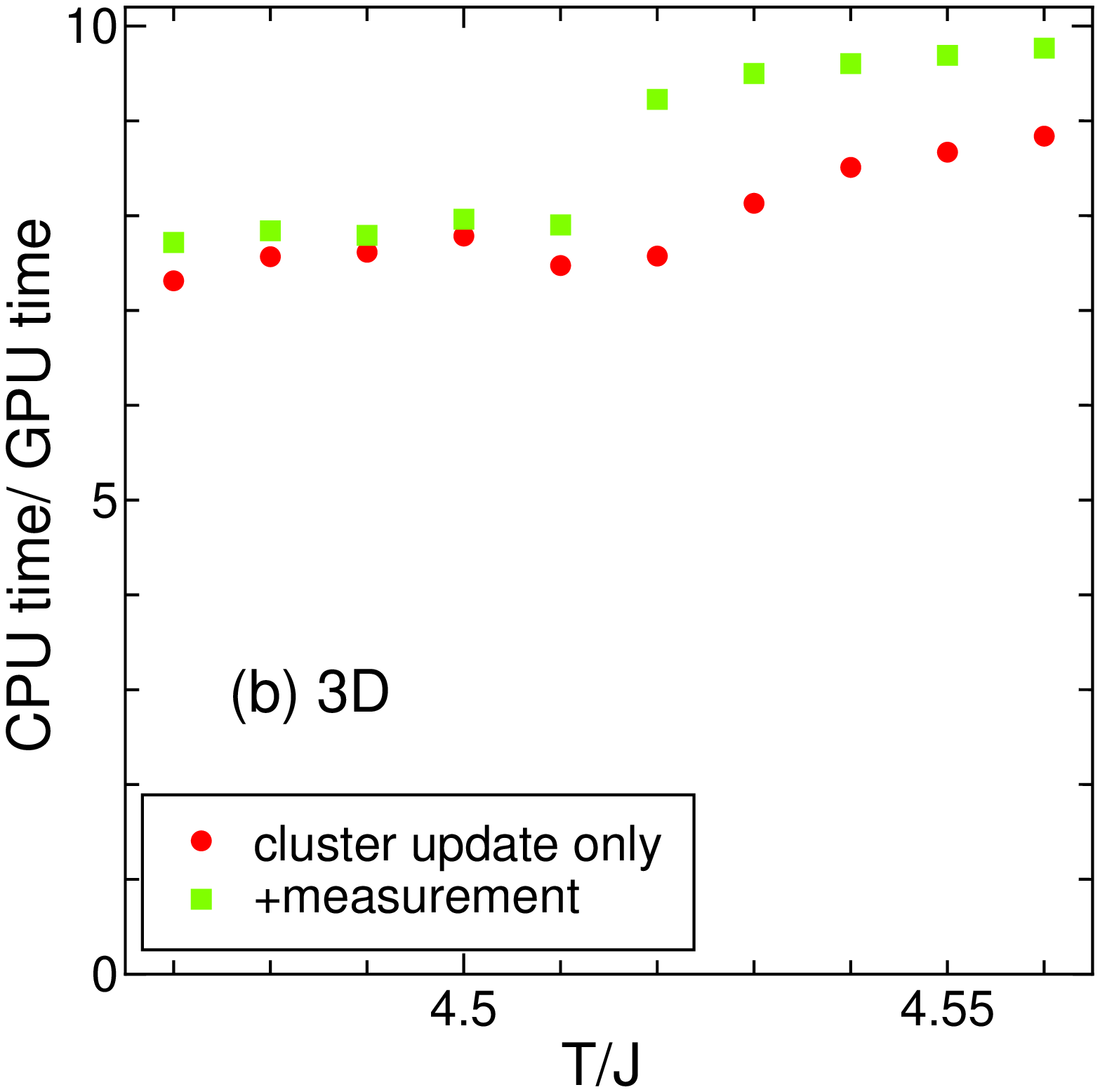}
\caption{(a) Temperature dependence of the acceleration rate for GPU computation for the 2D Ising model 
with $L=4096$ and (b) that for the 3D Ising model with $L=256$.}
\end{center}
\end{figure}

As an illustration, we plot the Binder ratio 
\begin{equation}
 U(T) = 1-\frac{<M(T)^4>}{3<M(T)^2>^2}
\end{equation}
of the 2D and 3D Ising model in figures 2(a) and (b), respectively. 
For the 2D Ising model, we discarded the first 5,000 single-cluster updates and 
the next 50,000 single-cluster updates were used for measurement.
The first 10,000 single-cluster updates were discarded and the next 100,000 single-cluster 
were used for measurement for the 3D Ising model. 

\begin{figure}
\begin{center}
\includegraphics[width=0.4\linewidth]{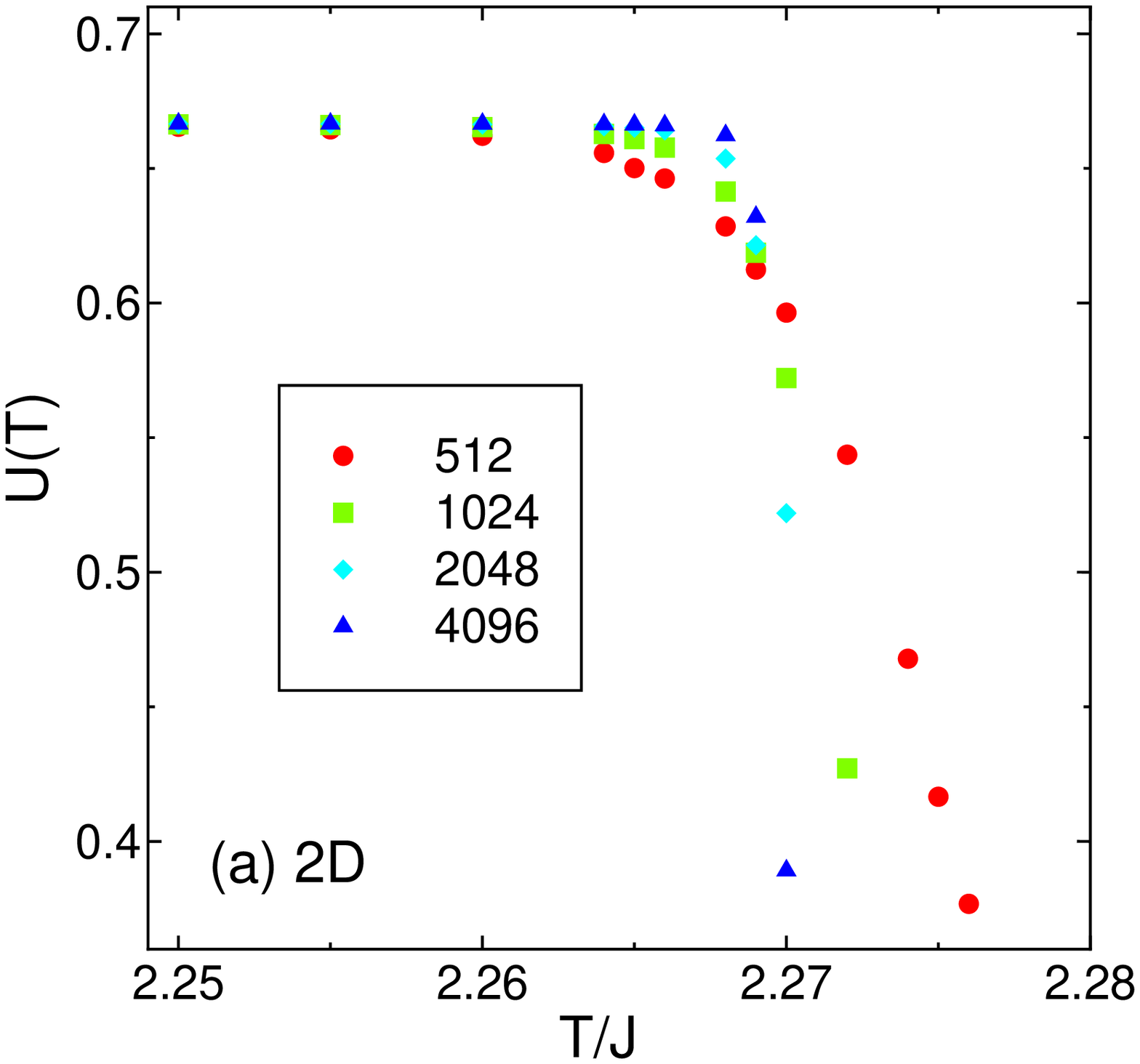}
\hspace{0.02\linewidth}
\includegraphics[width=0.4\linewidth]{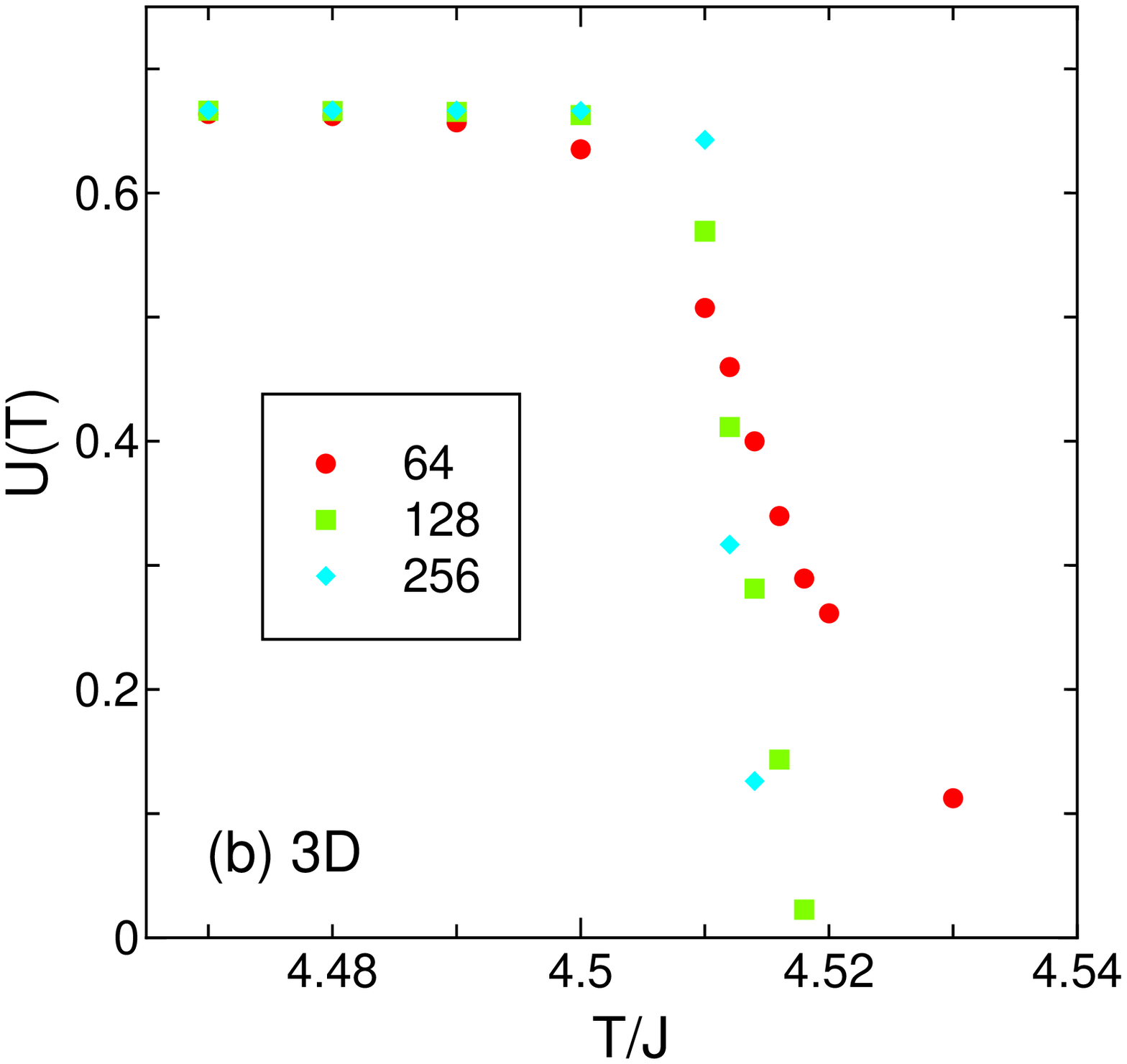}
\caption{(a) Binder ratio of the 2D Ising model for $L$=512, 1024, 2048 and 4096 and 
(b) that of the 3D Ising model for $L$=64, 128 and 256.}
\end{center}
\end{figure}

\section{Summary and discussion}
We have formulated a GPU parallel computing of the Wolff single-cluster 
Monte Carlo algorithm for the Ising model.  
We have compared the GPU computational time with the CPU computational time 
for the 2D and 3D Ising model. 
The GPU computational speed for the 2D Ising model with 
$L=4096$ is 5.60 times as fast as the CPU computational speed at the critical temperature, 
and that for the 3D Ising model with 
$L=256$ is 7.90 times as fast as the CPU computational speed. 
This acceleration rate is much higher than that reported by Kaupuzs {\it et al.} \cite{kaupuzs}. 

We have shown that the computational speed is accelerated for GPU 
over a wide range of temperatures for larger system sizes.  
The extension of this calculation to the Potts model is straightforward.  
It is also easily extended to continuous spin models, 
such as the classical XY model, using the Wolff's idea of 
embedded cluster \cite{wolff89}.  The application to 
quantum cluster algorithm \cite{evertz,kawashima,beard} is highly desirable. 
It is also interesting to apply this parallel calculation to 
the probability-changing cluster algorithm \cite{tomita}.

It is apparent from table \ref{tb:GPU_CPU_time_2D} and 
\ref{tb:GPU_CPU_time_3D} that the calculation with CPU is better for 
smaller lattices.  For the calculation of larger lattices, 
a hybrid calculation using CPU and GPU could be possible; 
that is, one starts with the calculation using CPU, and then switches 
to the calculation using GPU when the size of growing cluster 
exceeds some value.  In our code, one checks $\verb"(ic-in)"$, 
the number of wave-front spins. 
However, in such a hybrid calculation one needs the transfer of 
data between CPU and GPU.  We tested the performance 
of such a hybrid calculation.  In table \ref{tb:hybrid_2D}, 
we compare the average computational time at $T_c$ 
of the 2D Ising model with hybrid calculation, and that with 
only GPU and that with only CPU.
In the hybrid calculation we switch from CPU to GPU when 
the number of wave-front spins exceeds 100. 
For $L=1024$ in 2D, the trade-off 
between the saving time by the use of CPU for initial stage 
of growing cluster and the extra time of data transfer 
is compatible, that is, the total computational time is 
almost the same. However, for much larger lattices 
the parallelization in GPU becomes more prominent, 
and the time for data transfer becomes a bottleneck. 
Although a hybrid calculation is not so efficient in this 
algorithm, it will be interesting to pursue such an approach 
in other problems.

\begin{table*}[htbp]
\begin{center}
\begin{tabular}{rrrrr}
\hline
$L$ \quad & GPU \quad & hybrid \quad \quad & CPU \quad \\
\hline
  512 & 7.634 msec & 6.995 msec & 4.151 msec \\
 1024 & 15.80 msec & 16.89 msec & 20.97 msec \\
 2048 & 33.02 msec & 40.10 msec & 87.59 msec \\
 4096 & 81.37 msec & 111.8 msec & 412.0 msec \\
\hline
\end{tabular}
\caption{\label{tb:hybrid_2D}Comparison of average computational time 
per a single-cluster update at $T_c$ for the 2D Ising model.  
The time for hybrid calculation is compared with that with 
only GPU and that with only CPU.}
\end{center}
\end{table*}

We here mention the bottleneck of our method associated with the quasi-block
synchronization.  This does not mean that the quasi-block
synchronization itself takes long time,
but the latency is a problem when all blocks reach the quasi-block
synchronization point.
When a new spin is added to the cluster in the step (iii') of section 3,
our method needs to update a flag variable in global memory
to check whether that site belongs to the cluster or not.
Since the sites of the newly added spins are chosen randomly,
this task is not a coalesced memory access in global memory,
and the amount of the task becomes quite different for each block.
However, all blocks must wait for other blocks at the quasi-block
synchronization point. Thus, the latency becomes a bottleneck.
If we could equalize the load of the global memory access of each block,
the performance will be improved.  But such a trial will be a difficult job. 

We here mention about the GPU occupancy, which is available 
by the CUDA Visual Profiler.  The value of the GPU occupancy 
of our algorithm is 0.25, which is lower than 1. 
It is because we fix the number of threads in a block as 32. 
We could bring the GPU occupancy close to 1 by increasing 
the number of parallelization, but in our algorithm 
it also causes extra computational cost due to the extra warp 
together with that in the process of summing up 
the total number of newly added spins. 

After we finished the present work, we came to know that 
Hawick {\it et al.} \cite{hawick} also studied the CUDA 
implementation of the Wolff algorithm. 
They used a modified Connected Component Labelling 
for the assignment of the cluster.  Since the GPU performance 
of the Wolff update is not so good as that of 
the Metropolis update, they put more emphasis on 
the hybrid implementation of Metropolis and Wolff updates and 
the optimal choice of the ratio of both updates. 
It is interesting to compare their GPU implementation 
of the Wolff single-cluster algorithm with that of ours, 
which will be left to a future work. 

We finally make a comment on synchronization.  In CUDA, the function for 
inter-block synchronization is not natively provided.  
We have proposed an idea of 
a quasi-block synchronization with CUDA, which is effective 
for our single-cluster algorithm. 
This quasi-block synchronization algorithm can be used not only 
for a single-cluster 
Monte Carlo algorithm but also in many fields where the synchronizations of all threads 
in all blocks are required.

\section*{Acknowledgment}
This work was supported by a Grant-in-Aid for Scientific Research from
the Japan Society for the Promotion of Science.

\end{document}